\documentclass[rsi,reprint]{revtex4-1}
\usepackage{lineno,hyperref}
\usepackage[utf8]{inputenc}
\usepackage{blindtext, amsmath, amssymb,multirow,graphicx,url,subfig,float,epsfig,wasysym,colortbl}
\usepackage[utf8]{inputenc}
\usepackage{relsize}
\usepackage[font=small]{caption}
\usepackage[export]{adjustbox}
\usepackage[all]{hypcap}
\usepackage{dcolumn}
\usepackage{bm}
\modulolinenumbers[5]
\usepackage[T1]{fontenc}
\usepackage{mathptmx}
\begin{document}


\title{Energy Resolved Neutron Imaging for Strain Reconstruction using the Finite Element Method} 

\author{R. Aggarwal}
\email{Riya.Aggarwal@newcastle.edu.au}
\altaffiliation{School of Engineering, The University of Newcastle,\\
Newcastle, NSW 2308, Australia}
\author{M. H. Meylan}%
\affiliation{ 
School of Mathematics and Physical Sciences, The University of Newcastle,\\
Newcastle, NSW 2308, Australia
}
\author{B. P. Lamichhane}
\affiliation{%
School of Mathematics and Physical Sciences, The University of Newcastle,\\
Newcastle, NSW 2308, Australia
}%
\author{C M Wensrich}
\affiliation{School of Engineering, The University of Newcastle,\\
Newcastle, NSW 2308, Australia}


\date{\today}

\begin{abstract}
	A pulsed neutron imaging technique is used to reconstruct the residual strain within a polycrystalline material from Bragg edge strain images. This technique offers the possibility of a nondestructive analysis of strain fields with a high spatial resolution. A finite element approach is used to reconstruct the strain using a least square method constrained by the conditions of equilibrium. The procedure is developed and verified by validating for a cantilevered beam problem. It is subsequently demonstrated by reconstructing the strain from experimental data for a ring-and-plug sample, measured at the spallation neutron source \textsc{raden} at \textsc{j-parc} in Japan. The reconstruction is validated by comparison with conventional constant wavelength strain measurements on the \textsc{kowari} diffractometer at \textsc{ansto} in Australia.  It is also shown that the addition of a simple Tikhonov regularization can improve the reconstruction.
\end{abstract}


\maketitle 


\section{Introduction}
	\label{sec:intro}
	Energy resolved transmission imaging using time-of-flight spectroscopy of pulsed neutrons can give high wavelength--resolution Bragg edge transmission spectra of polycrystalline materials \cite{braggedge1,review,detector}. In these experiments, the term Bragg edge refers to a sudden increase in the relative transmission of a neutron beam passing through polycrystalline solids as a function of wavelength. A neutron, of wavelength $\lambda$, can be coherently scattered by crystal planes with lattice spacing $d$, provided that the scattering angle $\theta$ satisfies Bragg's law ($\lambda = 2d \sin \theta$). A sudden increase in transmission occurs once $\lambda = 2d$ is exceeded as a neutron cannot be scattered by more than $180^\circ$ \cite{review}, so neutrons are backscattered and no further diffraction occurs from that particular plane \cite{kisi2012applications,fitzpatrick2003analysis}. \\\\
	While other approaches exist \cite{braggedge1,braggedge2}, the process of measuring Bragg edges we use here relies on the measurement of the transmission spectra using the time-of-flight or energy-resolvedd techniques. This method requires a pulsed neutron source. Such a pulsed neutron source can be found in Japan (\textsc{j-parc}) \cite{raden,jparc}, \textsc{uk} (\textsc{isis}), and \textsc{usa} (\textsc{sns}). The greatest advantage of neutron strain tomography is that the incident beam flux is fully utilised, helping to reduce the data collection time. Modern technology uses a pixelated detector consisting of an array of up to $512\times512$ pixels with spatial resolution as small as $55 \mu m$ \cite{tof}. Such strain imaging raises the prospect of strain tomography, and several attempts have been made to solve the resulting tensor reconstruction problem over the past decade. These attempts have revolved around several special cases including axial-symmetry \cite{Kirkwood2015,Gregg2017TomographicRO}.\\\\
	As it stands, the inverse problem is ill-posed, and reconstruction of the strain is not possible without imposing further conditions \cite{inverse_prob}. Despite this problem, the reconstruction of the strain field under further assumptions has proven possible \cite{Kirkwood2015},\cite{chrisgranular}. Recently, it has been shown that by applying the condition of either equilibrium or compatibility, reconstruction is possible \cite{Abbey, Bragg_edge}. The main difference between these works is the choice of basis functions which represent the strain field.\\\\
	In this paper, we describe a method by which it is possible to tomographically reconstruct the elastic strain from a series of Bragg edge strain measurements using a finite element discretisation constrained by equilibrium. The proposed algorithm is tested on a cantilevered beam simulated data in two dimensions. It is shown to be capable of reconstructing a strain tensor field after imposing the equilibrium conditions \cite{PhysRevApplied.10.064034,Gauss_prob}. The algorithm is then applied to experimental data for a ring-and-plug geometry. We introduce a smoothing function to the minimisation problem with a regularisation parameter. Hence, minimising the value of the objective function will give us a regularised resistivity update equation to reduce the noise in the reconstructed images.
	
	\section{Longitudinal Ray Transform}
\begin{figure}[H]
		\centering
		\includegraphics[width=0.5\linewidth]{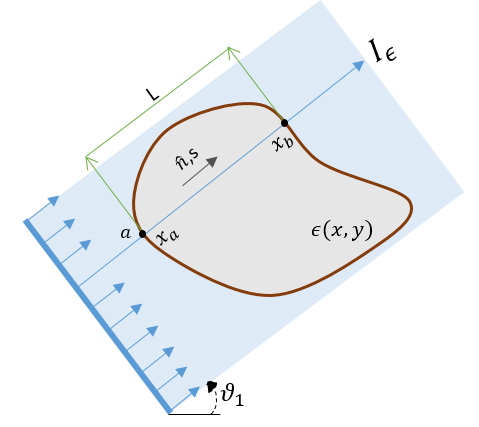}
		\caption{ Two-dimensional representation of Longitudinal Ray Transform: showing a ray entering the sample of the thickness $L$ at the position $a$ in the direction of $\hat{n}$. }
		\label{fig:LRT}
	\end{figure}

	We outline here the experimental technique which has recently been developed that provides information on the average strain component in the direction of the incident beam \cite{braggedge1}. As mentioned earlier, Bragg edges are formed by backscattering radiation. Hence, relative shifts in their position provide a measure of the average normal strain within a sample in the direction of the beam, i.e.,
	\begin{equation}
	\epsilon = \dfrac{\lambda-\lambda_0}{\lambda_0}.
	\end{equation}
	Therefore, the average strain within a body as measured by Bragg edge neutron transmission can be idealised as a line integral known as Longitudinal ray transform (LRT) which captures the average component of strain along the line $s$ in the direction of the unit normal $\hat{\mathbf{n}} = (n_i,n_j)^T=(\cos\vartheta,\sin\vartheta)^T$. We define
	\begin{equation}
	I_\epsilon(a,\vartheta)=\frac{1}{L}\int_{0}^{L} n_i \,\epsilon_{ij}(x(s,a),y(s,a))\, n_j \,ds,
	\label{LRTeqn}
	\end{equation}
	where $\epsilon_{ij}$ is the component of tensor strain field $\epsilon \in \mathbf{R}^{2\times2}$ which is mapped to an average normal component of a strain in the direction of $\hat{\mathbf{n}}$. The ray enters the sample at the position $x_a$ and $L$ is the ray length inside the sample for a particular angle $\vartheta$. This configuration is shown in Figure~\ref{fig:LRT}. This technique relies on the overall change in the lattice spacing along the ray \cite{Abbey,inverse_prob}. Measurements are taken in each orientation, $\vartheta_i$, where a profile is measured of the form $I_\epsilon(a,\vartheta_i)$. Inherent symmetry of the transform implies $180^\circ$ are sufficient; however, in practice, measurements are taken over an entire revolution, i.e., $360^\circ$. 
	Lionheart and Withers \cite{inverse_prob} demonstrated that the integral line LRT is a non-injective map from $\epsilon \to l_\epsilon(a,\vartheta)$ and hence the strain field produced by any given set of projection is not unique \cite{Sharafutdinov1994INTEGRALGO}. As a consequence, it is not possible to reconstruct the strain distribution within a body in the general setting. Hence, additional information (equilibrium or compatibility constraints) is required to make sure it is the required strain field from all the possibilities. To this end, several prior approaches have been developed that rely upon assumptions of compatibility or equilibrium to further constrain the problem. Compatible strain fields are those that can be written as the gradient of a displacement field in a simply connected body (i.e., conservative strain fields.). For Example, Abbey et al. \cite{Abbey} developed an algorithm using different basis functions along with compatibility constraints \cite{Bragg_edge}. Special cases have been considered including axis-symmetric systems \cite{Abbey,Abbey12,Gregg2017TomographicRO,Kirkwood2015} and granular systems \cite{chrisgranular} with equilibrium conditions. Unknown strain can also be reconstructed by using a machine learning technique known as Gaussian Process \cite{Gauss_prob, Gaussian} where equilibrium is used as a central technique to make sure strain is chosen uniquely. Also, arbitrary strain fields produced due to in-situ loadings have been reconstructed by using compatibility \cite{Chris2016,Bragg_edge}. We present here a reconstruction using the finite element method, noting that the finite element method has proved to be a very widely applicable method.
	
	\section{Solution using Finite Element Basis Functions}
\begin{figure}[H]
		\centering
		\includegraphics[width=0.5\linewidth]{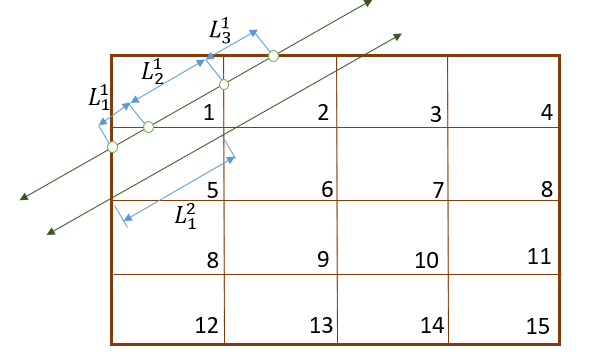}
		\caption{Finite element discretisation}
		\label{FemLrt}
	\end{figure}

	In our numerical implementation, each component of strain is approximated by a linear combination of basis functions, which come from the finite element method. The line integral~\eqref{LRTeqn} is solved in terms of the unknown strains which are equated to the Bragg edge measurements. The uniqueness of the solution is guaranteed by the equilibrium equation, which is imposed on the minimization problem used to calculate the backward map for the strain, in the form of extra constraints.
	We formulate the problem as follows:
	\[\epsilon(x,y) = \begin{bmatrix}
	\epsilon_{11}(x,y) & \epsilon_{12}(x,y) \\
	\epsilon_{21}(x,y) & \epsilon_{22}(x,y)
	\end{bmatrix} \]
	is the symmetric strain tensor field, i.e., $\epsilon_{12}(x,y)=\epsilon_{21}(x,y)$,
	\[\hat{\mathbf{n}}=\dfrac{\mathbf{x}_b - \mathbf{x}_a}{\|\mathbf{x}_b - \mathbf{x}_a \| } = (n_1,n_2),\] is normal component, and $L=$ length of a ray inside the sample/geometry. The main problem is to find the ray transform of the components of strain aligned with the direction of projection $\hat{\mathbf{n}} $ defined in equation \eqref{LRTeqn} rewritten in the form:
	\begin{equation} \label{LRT2}
	I_\epsilon = \dfrac{1}{L}\int_{0}^{L}\sum_{i,j=1,2}\,[ \,\epsilon_{ij}(a+s\hat{\mathbf{n}}) \,n_i\,n_j ]\,ds,
	\end{equation}
	where $\mathbf{x}_a=(x_a,y_a)$ and $\mathbf{x}_b=(x_b,y_b)$ are the entry and exit points of the ray respectively.
	The computational solution of the integral equation \eqref{LRT2} requires discretisation, i.e., the integral is expressed in terms of finitely many unknowns. We discretise the sample by using a quadrilateral mesh with $m$ nodes and $P$ elements. A visualisation of such a field over a rectangular sample discretised into rectangles is shown in the Figure~\ref{FemLrt} . A given ray can enter and exit the sample at arbitrary points. Hence, by applying discretisation we obtain the approximate problem as follows:
	\begin{equation}
	I_\epsilon \approx \dfrac{1}{L}\sum_{P} \int_{0}^{\triangle L_P} \hat{\mathbf{n}}^T \epsilon^P_{ij}\, \hat{\mathbf{n}}\, ds,
	\label{eqn:integral}
	\end{equation}
	where $P=\{P_1,P_2, \ldots , P_n\}$ is the set of elements and $\triangle L_P$ is the length of the ray in each element. Note that this length will be zero in many of the elements. An example of such a discretisation is shown in Figure~\ref{FemLrt} where the first ray is intersecting with the elements $1,2 \,\, \text{and}\,\, 5$, whereas the second ray is intersecting the elements $2,3,5 \,\,\text{and}\,\, 6$. The strain in any element depends on the strain value at the corner of the quadrilaterals which are the unknowns. In general each node strain will influence the strain in four elements. The strain at each node has three components.
	\noindent The strain is expressed using the standard basis function as follows:
		\begin{equation}
		\epsilon_{ij}^P(x,y) = \beta_{ij}^P+\gamma_{ij}^P\,x+\eta_{ij}^P\,y+\zeta_{ij}^P\,x\,y\,\,,\,\, i,j=\{1,2\},
		\label{elem_eqn}
		\end{equation}
		where, $\beta_{ij}^P,\gamma_{ij}^P$, $\eta_{ij}^P$ and $\zeta_{ij}^P$ are the coefficients which are determined from the nodal values of strain in the element $P$.
		The measurement for the first ray in Figure~\ref{FemLrt} can be approximated by:
	\begin{equation}
	I_\epsilon \approx \dfrac{1}{L}\Bigg[\int_{0}^{\triangle L_1} \hat{\mathbf{n}}^T \epsilon_{ij}^1\, \hat{\mathbf{n}}\, ds+\int_{0}^{\triangle L_2} \hat{\mathbf{n}}^T \epsilon_{ij}^2\, \hat{\mathbf{n}}\, ds+\int_{0}^{\triangle L_5} \hat{\mathbf{n}}^T \epsilon_{ij}^5\, \hat{\mathbf{n}}\, ds\Bigg],
	\end{equation}
	where $L=\displaystyle\sum_{P} L_P$, $L_P$ is the segment of the ray inside the $P^{th}$ rectangle where $P=1, 2$ and $5$ and $\epsilon_{ij}^P$ is strain value in the element $P$. The value of
	$\beta_{ij}^P$, $\gamma_{ij}^P$, $\eta_{ij}^P$ and $\zeta_{ij}^P$ for each element can be found by solving the following equation
	\begin{equation} \begin{bmatrix}
	\epsilon_{ij}^P (x^1,y^1)\\ \epsilon_{ij}^P (x^2,y^2)\\ \epsilon_{ij}^P (x^3,y^3)\\ \epsilon_{ij}^P (x^4,y^4)
	\end{bmatrix}= \begin{bmatrix}
	1 & x^1 & y^1& x^1y^1\\
	1 & x^2 & y^1& x^2y^2\\
	1 & x^3 & y^3& x^3y^3\\
	1 & x^4 & y^4& x^4y^4\\
	\end{bmatrix} \begin{bmatrix}
	\beta_{ij}^P\\
	\gamma_{ij}^P\\
	\eta_{ij}^P\\
	\zeta_{ij}^P\\
	\end{bmatrix}
	\label{strain_node mat}
	\end{equation}
	where $x^{i}$ and $y^{i}$ are the coordinates of the corner points of the quadrilaterals.
	In general, each element $P$ will have different ray entry and exit points.
	Using the above line integral expression and evaluating the basis functions for each element, the integral can be reformulated in terms of nodal strain and we obtain a system of equations of the form
	$$\begin{bmatrix}
	I_1 \\
	I_2\\
	\vdots\\
	I_{N-1}\\
	I_{N}\\
	\end{bmatrix}= \mathbf{K}\begin{bmatrix}
	\epsilon_{11}^1 \\\vdots\\\epsilon_{22}^{m} \\
	\end{bmatrix},$$
	where $N$ is the total number of projections, $I_v$, $v=\{1,...,N\}$, is the value from each measurement, $\mathbf{K}$ is the matrix derived from the integrals in equation~\eqref{eqn:integral} expressed in terms of the nodal strain through equation~\eqref{strain_node mat}. We can write this in compact form as
	\begin{equation*}
	\mathbf{I}= \textbf{K}\,\, \boldsymbol{\epsilon}, 
	\end{equation*}
	where $\mathbf{I}$ is a vector containing all of the Bragg edge strain measurements, $\mathbf{K}$ is the coefficient matrix with elements that contain unit direction vector components and shape function evaluations which will be a sparse matrix, and $\boldsymbol{\epsilon} = \begin{bmatrix}
	\epsilon_{11}^1 &\cdots&\epsilon_{22}^{m} \\
	\end{bmatrix}^T$, $m =\{ \text{number of nodes}\}$, is a vector containing all the unknowns for each element.
	Once the matrix $\mathbf{K}$ and vector $\mathbf{I}$ have been formed, the problem is reduced to one of solving the linear algebraic system of equations for the unknown coefficients represented by vector $\boldsymbol{\epsilon}$. In practice the system is usually over-determined since the number of unknown coefficients is relatively small compared to the amount of experimental data available. Furthermore, the $\mathbf{K}$
	matrix will be sparse. \\\\
	As it was pointed out before by Lionheart and Withers \cite{inverse_prob} the strain field is not uniquely defined within an object from these measurements. For this reason with apply the constraints to our problem obtained by solving the equilibrium equations. From Hooke's law, the equilibrium equation can directly be written in terms of strain. Assuming plane stress condition, the equilibrium conditions can be written as:
	\begin{equation}
	\frac{\partial}{\partial x}(\epsilon_{11}+\nu \epsilon_{22})+\frac{\partial}{\partial y}(1-\nu)\epsilon_{12}=0,
	\label{equil1}
	\end{equation}
	\begin{equation}\frac{\partial}{\partial y}(\epsilon_{22}+\nu \epsilon_{11})+\frac{\partial}{\partial x}(1-\nu)\epsilon_{12}=0,
	\label{equil2}
	\end{equation}
	where, $\nu$ is the Poisson's ratio. To reconstruct the strain, the equilibrium equations \eqref{equil1} and \eqref{equil2} are integrated over each element $P$, which will lead us to the following using \eqref{elem_eqn}
	$$\iint_P\Big[\gamma_{11}^P+\zeta_{11}^Py+\nu(\gamma_{22}^P+\zeta_{22}^Py)+(1-\nu)(\eta_{12}^P+\zeta_{12}^Px)\Big]dx dy=0, $$
	$$\displaystyle\iint_P\Big[\eta_{22}^P+\zeta_{22}^Px+\nu(\eta_{11}^P+\zeta_{11}^Px)+(1-\nu)(\gamma_{12}^P+\zeta_{12}^Py)\Big]dx dy=0.$$
	This provides another set of a system of equations
	\begin{equation}
	\mathbf{C} \boldsymbol{\epsilon}=0,
	\end{equation}
	where $\mathbf{C}$ represents the equilibrium integral matrix, which has two rows.
	Solutions to the minimisation problem were found by least-square fitting \cite{lsqr},
	where the problem is reduced to: find a vector $\boldsymbol{\epsilon}$ such that
	\begin{equation}
	\min_{\mathbf{C} \boldsymbol{\epsilon} = 0} \|\mathbf{K} \boldsymbol{\epsilon}-\mathbf{I}\|_2,
	\label{minimisation}
	\end{equation}
	The minimisation problem \eqref{minimisation} is solved straightforwardly using least squares.
	
	\section{Cantilevered Beam}
\begin{figure}[H]
		\centering
		\includegraphics[width=0.5\linewidth]{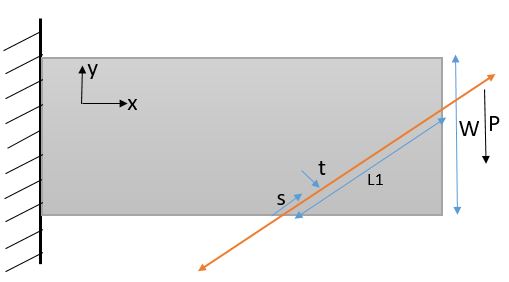}
		\caption{Cantilevered Beam Geometry with $\text{beam length }(L) = 20mm, \text{width}~(W)=10mm, \text{thickness}~ (t)=5mm, E=200GPa,\nu=0.3,I=tW^3/12$. }
		\label{Beamgeometry}
	\end{figure}

\begin{figure}[H]
		\centering
		\includegraphics[width=0.95\linewidth,frame]{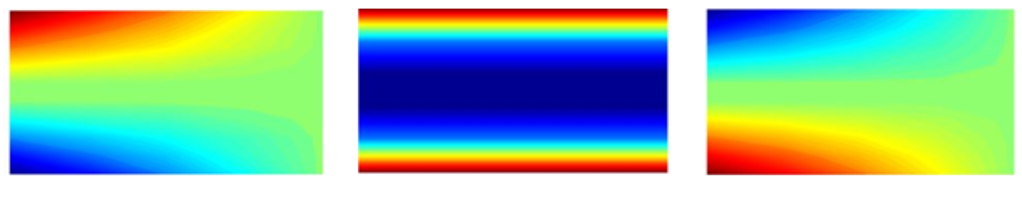}
		\includegraphics[width=0.95\linewidth,frame]{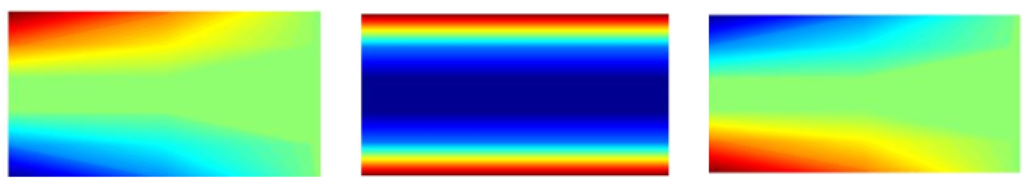}
		\caption{Beam solution (Top figure: Reconstructed solution $\epsilon_{xx},\epsilon_{xy},\epsilon_{yy}$ and lower figure: True solution $\epsilon_{xx},\epsilon_{xy},\epsilon_{yy}$)}
		\label{beamsolution}
	\end{figure}

	To demonstrate the performance of the proposed algorithm, a well-known 2D cantilevered beam problem is studied, which was previously examined by Wensrich et al. \cite{Chris2016}. We consider the 2D strain field for beam geometry of the rectangle $[0,12] \times [0,10]$ with the load $P$ of 2 kN displayed in Figure~\ref{Beamgeometry}. Material properties of the beam are representative of common steel, whereas other parameters are mentioned below. This beam problem is excellent for testing the algorithm since the analytical solutions to the strain field exist.
	\begin{flushleft} Assuming plane stress, the Saint-Venant approximation to the strain field is \cite{Beamprob}:
		$$\epsilon(x,y)=
		\begin{bmatrix}
		(L-x)y & -\frac{(1+\nu)}{2}((\frac{W}{2})^2 - y^2)\\
		-\frac{(1+\nu)}{2}((\frac{W}{2})^2 - y^2) & -\nu (L-x)y \\
		\end{bmatrix} \frac{P}{EI}.$$\end{flushleft}
	This is shown in Figure~\ref{beamsolution}. A finite element model of the system was constructed, with a rectangular mesh. The reconstructed solution to the strain field for the cantilevered beam is shown in the Figure~\ref{beamsolution}.
	\noindent We found that the proposed reconstruction algorithm is extremely effective in achieving strain field reconstruction. A finite element model of the system was constructed, with a structured quadrilateral mesh size $4\times4$. Simulation results suggest that the reconstruction algorithm can converge to an adequate reconstruction provided that measurements are taken over the entire $360^\circ$ of a sample. Problem discretisation and numerical errors can undoubtedly contribute to an imperfect reconstruction (with more noise). Rapid convergence to the true solution was observed as the number of projections was increased. This convergence provides confidence in the ability of the algorithm to converge to a true solution in the presence of real experimental uncertainties.
	
	\section{Reconstruction of the Offset Ring-and-Plug}\label{sec4}
\begin{figure}[H]
		\centering
		\includegraphics[width=0.5\linewidth]{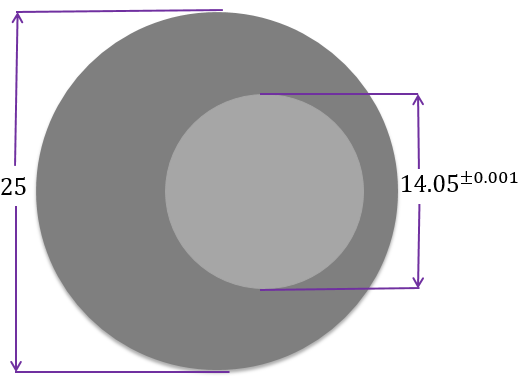}
		\caption{Sample Geometry (all dimensions are in mm.)}
		\label{RP}
	\end{figure}

\begin{figure}[H]
		\centering
		\subfloat[MT1]{\includegraphics[width=0.4\linewidth]{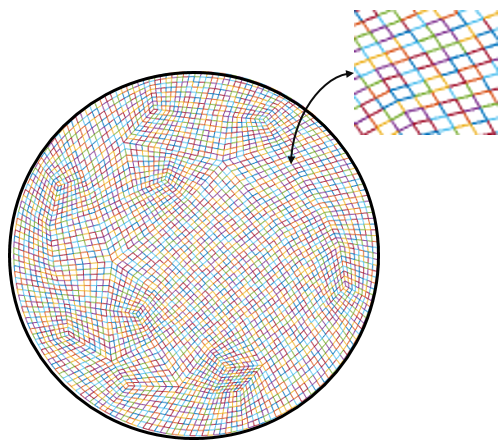}} \quad
		\subfloat[MT2]{\includegraphics[width=0.4\linewidth]{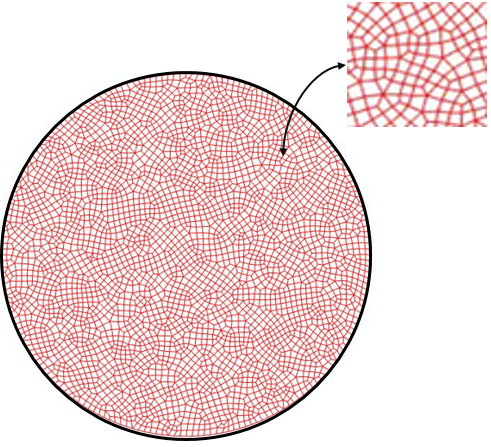}}
		\caption{Two types of Mesh Patches: (a) Structured Mesh and (b) Unstructured Mesh}
		\label{Meshpatches}
	\end{figure}

\begin{figure}[H]
		\includegraphics[width=0.3\linewidth]{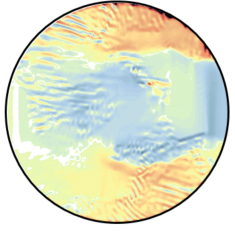}
		\includegraphics[width=0.3\linewidth]{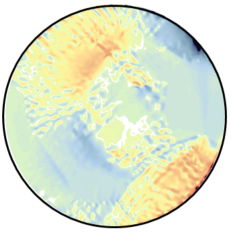}
		\includegraphics[width=0.3\linewidth]{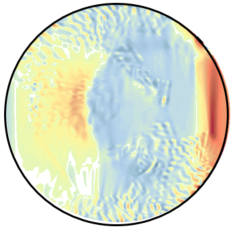}
		\includegraphics[width=0.06\linewidth]{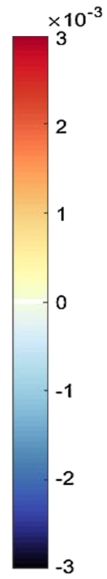}
		\caption{Ring-and-plug reconstructed strain field $\epsilon_{xx},\epsilon_{xy},\epsilon_{yy}$ for mesh type MT1}
		\label{solutionstructured}
	\end{figure}

\begin{figure}[H]
		\includegraphics[width=0.3\linewidth]{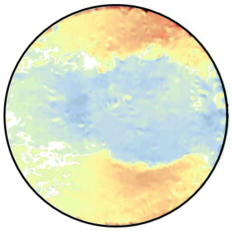}
		\includegraphics[width=0.3\linewidth]{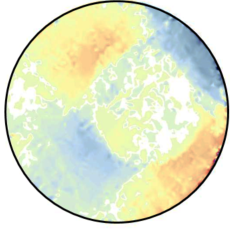}
		\includegraphics[width=0.3\linewidth]{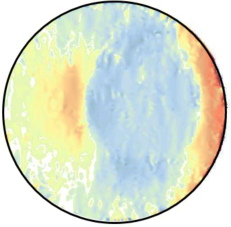}
		\includegraphics[width=0.06\linewidth]{bar1}
		\caption{Ring-and-plug reconstructed strain field $\epsilon_{xx},\epsilon_{xy},\epsilon_{yy}$ for unstructured mesh type MT2}
		\label{solutionunstructured}
	\end{figure}

	We now test the algorithm on experimental data for an offset ring-and-plug sample, which was used previously \cite{PhysRevApplied.10.064034}.
	The sample geometry of the offset ring-and-plug is shown in Figure~\ref{RP}. The described sample contained a total interference of $40 \pm 2\, \mu m$ produced through cylindrical grinding. More details about the sample can be found in \cite{PhysRevApplied.10.064034}. A steel bar EN26 was heated to relieve stress and provide a uniform structure prior to the assembly. The final hardness of the sample was 290 HV. The strain profile was measured on \textsc{raden} together with an MCP detector at a distance of 17.9 m from the source of the beam.
	\textsc{raden} is an energy--resolved neutron imaging instrument at the Japan Proton Accelerator Research Complex (\textsc{j-parc}), Japan, \cite{raden,jparc} was used to obtain the relative shifts of the Bragg edge corresponding to the lattice plane of the offset ring-and-plug steel sample. Neutron strain scanning was carried out on \textsc{kowari} a residual stress diffractometer at Australian Nuclear Society and Technology Organisation (\textsc{ansto}), Australia to provide independent validation of our reconstruction.
	A finite element quadrilateral mesh is used to discretise the domain of the sample, as shown in Figure~\ref{Meshpatches}. Two types of mesh patches have been considered: structured and unstructured meshes.
	Reconstruction results can be seen with each mesh type in Figure~\ref{solutionstructured} and \ref{solutionunstructured}. Again, as mentioned before for beam problem that the reconstructed strain field contains noise which can be from numerical discretisation or measurements. Some techniques are explained in the next section to cope up with noise present in the strain field.
	\begin{figure}[H]
		\includegraphics[width=0.3\linewidth]{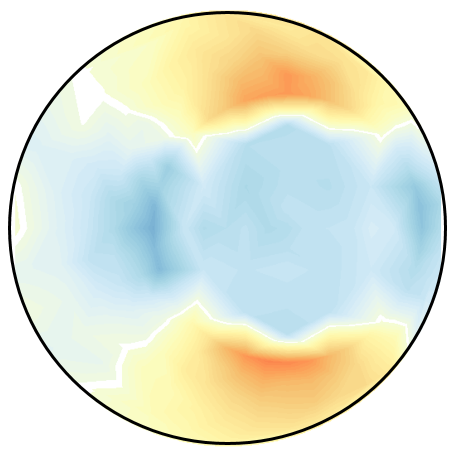}
		\includegraphics[width=0.3\linewidth]{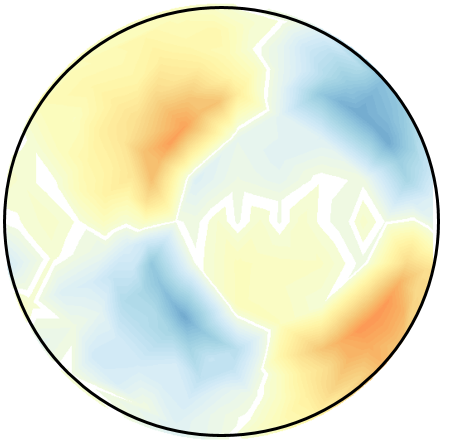}
		\includegraphics[width=0.3\linewidth]{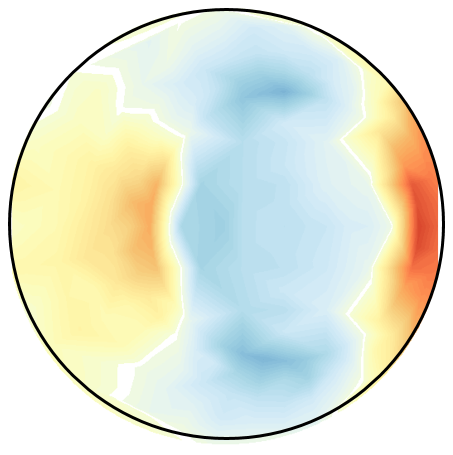}
		\includegraphics[width=0.06\linewidth]{bar1}
		\caption{Ring-and-plug strain images obtained from \textsc{kowari}}
		\label{Kowari}
	\end{figure}
	Different type of the mesh has been shown variation results, which proves that our algorithm is highly dependent on the mesh. It was observed that results with an unstructured mesh show better agreement than the structured mesh in terms of noise. This is because the unstructured mesh is evenly distributed throughout the sample domain, unlike the structured mesh, resulting in reduced numerical noise. Results are then compared with the pointwise measurement of strain on \textsc{kowari} - the constant wavelength diffractometer shown in Figure~\ref {Kowari} with the reconstructed transmitted measurements of strain on \textsc{raden} shown in  Figures~\ref{solutionstructured} and \ref{solutionunstructured} which looks promising.
	
	\section{Tikhonov Regularisation}
\begin{figure}[H]
		\includegraphics[width=0.3\linewidth]{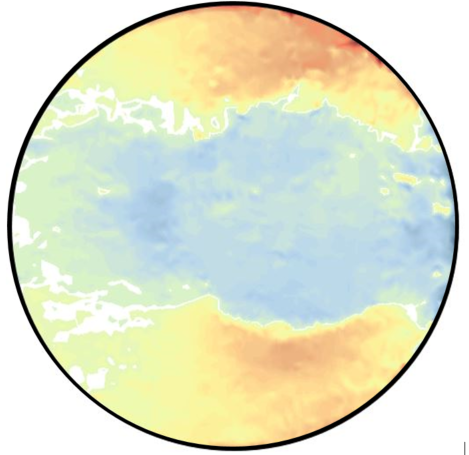}
		\includegraphics[width=0.3\linewidth]{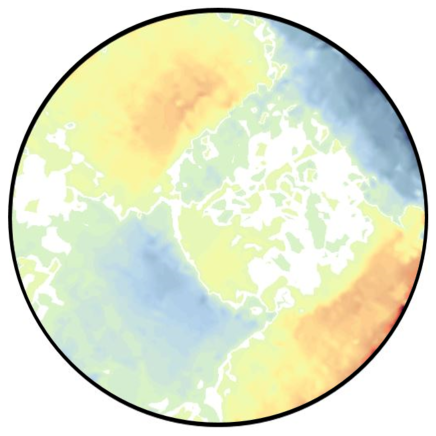}
		\includegraphics[width=0.3\linewidth]{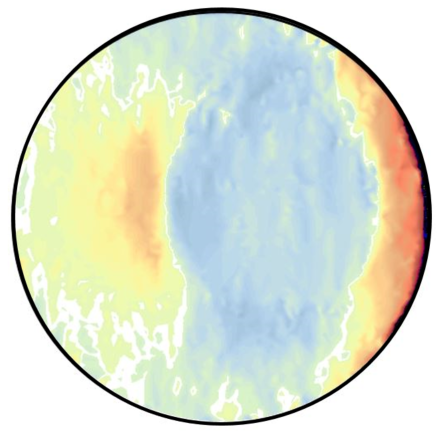}
		\includegraphics[width=0.055\linewidth]{bar1}
		\caption{Regularised strain field $\epsilon_{xx}\,,\epsilon_{xy},\,\epsilon_{yy}$ respectively for unstructured mesh type, with $S$ as stiffness matrix and $\alpha=0.001$}
		\label{tik_stiff1}
	\end{figure}

\begin{figure}[H]
		\includegraphics[width=0.3\linewidth]{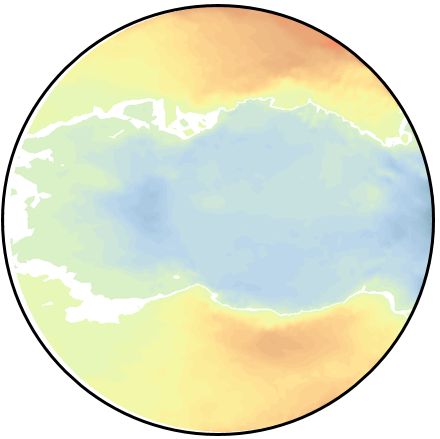}
		\includegraphics[width=0.3\linewidth]{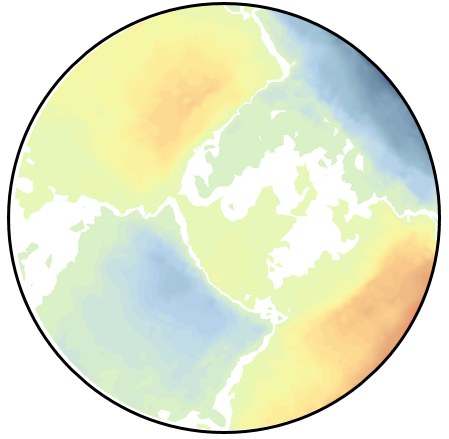}
		\includegraphics[width=0.3\linewidth]{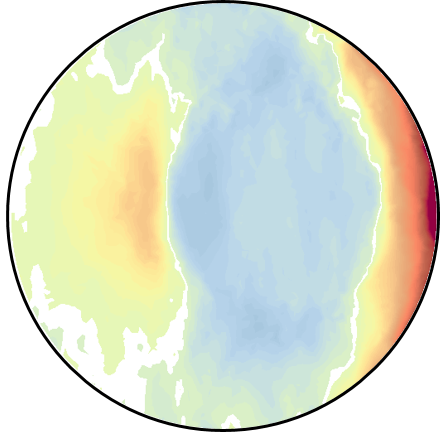}
		\includegraphics[width=0.06\linewidth]{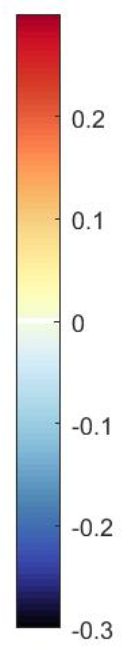}
		\caption{Regularised strain field $\epsilon_{xx}\,,\epsilon_{xy},\,\epsilon_{yy}$ respectively for unstructured mesh type with $S$ as stiffness matrix and $\alpha=0.005$}
		\label{tik_stiff2}
	\end{figure}

\begin{figure}[H]
		\includegraphics[width=0.3\linewidth]{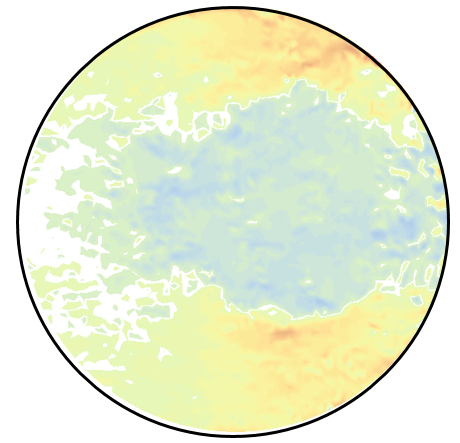}
		\includegraphics[width=0.3\linewidth]{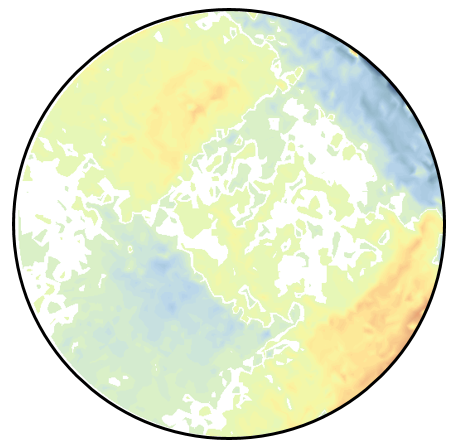}
		\includegraphics[width=0.3\linewidth]{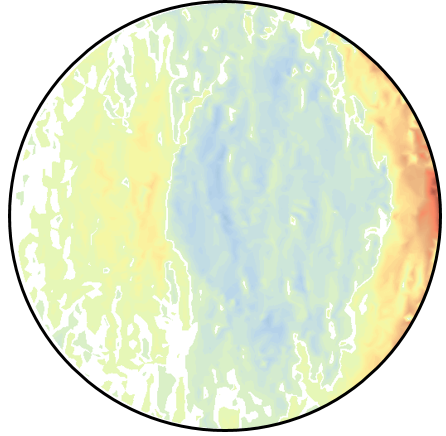}
		\includegraphics[width=0.06\linewidth]{bar1}
		\caption{Regularised strain field $\epsilon_{xx}\,,\epsilon_{xy},\,\epsilon_{yy}$ respectively for unstructured mesh type, with $S$ as identity matrix and $\alpha=0.005$}
		\label{tik_iden}
	\end{figure}

\begin{figure}[H]
		\includegraphics[width=0.3\linewidth]{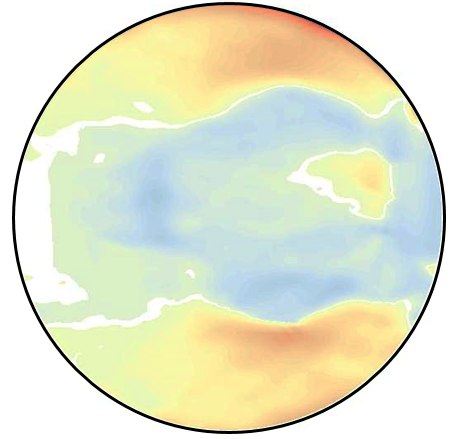}
		\includegraphics[width=0.3\linewidth]{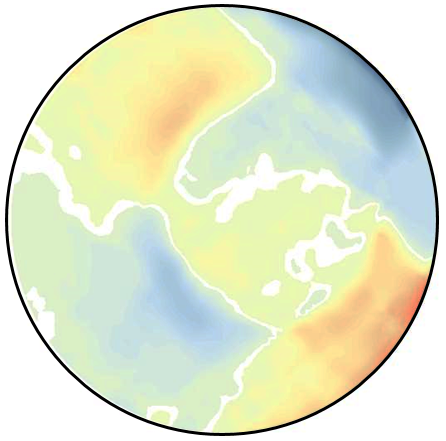}
		\includegraphics[width=0.3\linewidth]{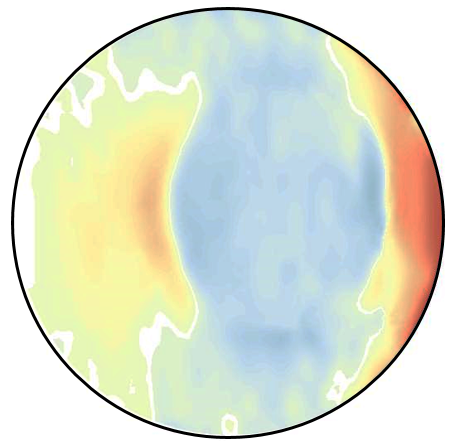}
		\includegraphics[width=0.06\linewidth]{bar1}
		\caption{Regularised strain field $\epsilon_{xx}\,,\epsilon_{xy},\,\epsilon_{yy}$ respectively for structured mesh type, with $S$ as stiffness matrix and $\alpha=0.005$.}
		\label{tik_stiff1_mesh1}
	\end{figure}

	Until this point our reconstruction does not involve any smoothing. As a result, reconstruction images have noise which can arise from different sources such as from instrumental measurement noise or the simulation procedure. To manage this noise, Tikhonov regularisation is used \cite{golub1999tikhonov}. The Tikhonov regularised estimate is defined as the solution of the following minimisation problem where the first term is the same euclidean norm used before in equation \eqref{minimisation}. The second term is known as the regulariser which captures the prior knowledge and behavior of $\boldsymbol{\epsilon}$ through an additional penalty term:
	\begin{equation*}
\boldsymbol{\epsilon}_R = \min\|\mathbf{K}\boldsymbol{\epsilon}-\mathbf{I}\|^2+\alpha \|B\boldsymbol{\epsilon}\|^2_2,
	\end{equation*}
	where $\| \cdot \|_2$ is Euclidean norm, $\alpha > 0$ is the regularisation parameter which specifies the amount of regularisation. The effect of regularisation can be varied due to the scale of the matrix B. The matrix B is a block diagonal matrix where the block diagonal entries can be chosen in several ways such as zero matrix which will bring our problem back to unregularised least square problem, it can be identity matrix shown in Figure~\ref{tik_iden}. Hence, in our case, B is chosen as the block diagonal matrix as
	$$B = \begin{bmatrix}
	S & 0& 0\\
	0& S& 0 \\
	0 & 0 & S
	\end{bmatrix} \in \mathbf{R}^{3n \times 3n}, $$ where $S_{ij}=\int_\Omega \triangledown \phi_i \cdot \triangledown \phi_j$ for $i,j=1,2,...,n$, and $\phi_i, \forall ~i,$ are the standard basis functions for quadrilateral.\\\\
	Numerically, the minimum is achieved by solving a linear least-square problem of the form:
	$$\boldsymbol{\epsilon}_R = \min \Bigg\|\begin{pmatrix}
	K \\
	\alpha B
	\end{pmatrix}\boldsymbol{\epsilon}-\begin{pmatrix}
	\mathbf{I}\,\\
	0
	\end{pmatrix} \Bigg\|^2_2 .$$
	Above equation is solved in \textsc{matlab} with a built-in function "lsqlin".
	The main problem here is to determine proper regularisation parameter $\alpha$; if the parameter is large, the solution will deviate from the true solution, and if the parameter is small then there will not be any significant difference in the noise. For now, we find this parameter by trial and error; however, it can also be achieved using optimisation methods \cite{wahba1990spline}.
	The effect of Tikhonov regularisation can be seen in the Figures~\ref{tik_stiff1} - \ref{tik_stiff1_mesh1}, where the difference is shown for different values of $\alpha$.

	
	\noindent To summarise, in Tikhonov regularisation, we approximate the minimum norm by least squares. Least square solution $\phi_R$ depends on $\mathbf{K}\phi$, by a vector depending on the regularisation parameter $\alpha >0$. Reconstruction is done on finite element rectangular mesh with 3688 elements for the unstructured mesh and 3776 elements for structured mesh type. Minimization problem for both cases (with and without regularisation) is solved in \textsc{matlab} by using built-in function 'lsqlin' which were then plotted by using 'scatteredInterpolant' with linear map fitting. The proposed algorithm does not solely depend on the sample geometry and hence, can be extended to three-dimensional sample bodies. The true difficulty will be the computational cost and time, since the size of the problem will be larger.
	
	\section{Conclusion}
	The proposed method helps to reconstruct entire strain field, satisfying equilibrium with no assumptions of compatibility and hence is suitable for reconstructing residual strain fields. The 
	\textsc{kowari} strain and \textsc{raden} reconstructed strain field measurements show closer agreement all together with regularisation. Feasibility of the proposed algorithm is done for the offset ring and plug problem, and results are compared to the other technique. The proposed method was validated with simulated data and strain estimates from experimentally measured data at \textsc{j-parc}, Japan, which were compared to the strain calculation from a conventional diffraction method obtained at \textsc{kowari}.
	In two dimensions, full strain fields tomography using Bragg edge images can now be achieved using physical constraints as equilibrium. This method opens up further research for future investigations, including extending this technique to three dimensions. Also, the proposed method allows us to use adaptive meshes which can focus on the highly strained area in the sample, which can be achieved by calculating gradient over the sample.
	\section{Data Availability Statement}
	Most of the data that supports the findings of this study are available within the article and rest of the data are available from the corresponding author upon reasonable request.
	\section{Acknowledgements}
	The Australian Research Council supports this work through a Discovery Project Grant. RA's scholarship is supported by ARC grant, UNIPRS, UNSR5050 Central School.
	
	\section{References}
	\bibliography{bib}


\end{document}